\documentclass[11pt,twocolumn]{article}

\pdfoutput=1

\usepackage[round]{natbib}   

\usepackage[english]{babel}

\usepackage{graphicx}
\usepackage{amsmath}

\usepackage{enumitem}

\usepackage{float}
\usepackage{ifthen}

\usepackage{hyperref}

\usepackage{titlesec}

\titlespacing*{\section}{0pt}{1.1\baselineskip}{\baselineskip}

\begin{document}

\def\hmath#1{\text{\scalebox{1.5}{$#1$}}}
\def\lmath#1{\text{\scalebox{1.4}{$#1$}}}
\def\mmath#1{\text{\scalebox{1.2}{$#1$}}}
\def\smath#1{\text{\scalebox{.8}{$#1$}}}

\def\hfrac#1#2{\hmath{\frac{#1}{#2}}}
\def\lfrac#1#2{\lmath{\frac{#1}{#2}}}
\def\mfrac#1#2{\mmath{\frac{#1}{#2}}}
\def\sfrac#1#2{\smath{\frac{#1}{#2}}}

\def\pow{^\mmath}



\twocolumn[

\begin{center}
{\bf \Large {
The first and second order approximations of the third-law}}
\vspace{2mm}
\\
{\bf \Large {
moist-air entropy potential temperature.}}
\\
\vspace*{3mm}

{\Large by Pascal Marquet$\:{}^{(1)}$}. \\
\vspace*{3mm}
{\large ${}^{(1)}$ M\'et\'eo-France CNRM/GMAP and CNRS UMR-3589.
 Toulouse. France.}
\\ \vspace*{2mm}
{\large  \it E-mail: pascal.marquet@meteo.fr}
\\ \vspace*{2mm}
{\large  Submitted to the}
{\large \it \underline{Monthly Weather Review}}
{\large  -- 2 March, 2019. 
 Revised on 19 May 2019.} 
\vspace*{1mm}
\end{center}

\begin{center}
{\large \bf Abstract}
\end{center}
\vspace*{-3mm}

\hspace*{7mm}
It is important to be able to calculate the moist-air entropy of the atmosphere with precision. 
A potential temperature has already been defined from the third law of thermodynamics for this purpose.
However, a doubt remains as to whether this entropy potential temperature can be represented with simple but accurate first- or second-order approximate formulas.
These approximations are rigorously defined in this paper using mathematical arguments and numerical adjustments to some datasets.
The differentials of these approximations lead to simple but accurate formulations for tendencies, gradients and turbulent fluxes of the moist-air entropy.
Several physical consequences based on these approximations are described and can serve to better understand moist-air processes (like turbulence or diabatic forcing) or properties of certain moist-air quantities (like the static energies).

\vspace*{4mm}

%
] 

\section{Introduction.} 
\label{section_intro}
\vspace*{-3mm}

The possibility of calculating the entropy of moist air can allow the study of its variations within the atmosphere, both in space and in time.
This should lead to a better understanding of the turbulent processes, as well as some renewal for other aspects of the energetics of the atmosphere.

To do so, the entropy $S$ of a moist-air parcel of mass $m$ can be computed by summing the partial entropies of dry air and water vapour, plus the partial entropies of possible liquid water and ice condensed species contained in clouds or in precipitations.
The quantity $s = S/m$ is the specific entropy defined per unit mass of moist air.
The specific value of moist-air entropy defined in \citet[hereafter HH87]{Hauf_Holler_87} is in full agreement with the third law of thermodynamics, with reference values for entropies defined at zero Kelvin for the more stable solid states of all atmospheric species.
The third-law entropy of moist air of HH87
is written in \citet[hereafter M11]{Marquet_2011} 
in terms of an entropy potential temperature $\theta_s$, leading to
\vspace{-1mm}
\begin{align}
 s  & \: = \; c_{pd} \: \ln(\theta_{s}) \: + \:  s_{ref}
  \label{eq_s_thetas}
  \: ,
\end{align}
where both 
$s_{ref}$ and the specific heat at constant pressure of dry air $c_{pd}$
are constant for the range of absolute temperature in the atmosphere (between $180$~K and $330$~K).

Equation~(\ref{eq_s_thetas}) means that $\theta_s$ becomes truly synonymous with the specific moist-air entropy ($s$), whatever the local thermodynamic properties of temperature, pressure and humidity.
This is a generalisation of the dry-air relationship $s \: = \; c_{pd} \: \ln(\theta) + s_0$ first derived by 
\citet{Bauer_1910},
in which the properties of the specific entropy of a given perfect gas (like the dry air) are not affected by the arbitrary constant of integration $s_0$.
The specific entropy is defined differently from (\ref{eq_s_thetas}) in HH87, where  the constant values of $c_{pd}$ and $s_{ref}$ are replaced by values that depend on the local water content, which prevents the potential temperature $\theta_s$ of HH87 from varying like entropy.

Although the entropy can be studied by itself, it is of common practice in meteorology to study  the properties of potential temperatures instead, like $\theta_s$.
However, the formulation for $\theta_s$
which comes from Eq.~(\ref{eq_s_thetas}) and which is recalled in section~\ref{section_thetas1} leads to the same degree of complexity as the complete formulations of \citet{Emanuel_94} for the liquid-water ($\theta_l$) and equivalent ($\theta_e$) potential temperatures.
These complete formulations are almost never used and only approximate formulations are considered, like the equivalent potential temperature of \citet{Betts_73}.
Therefore, it seems desirable to seek the first- and second-order approximations of the entropy potential temperature $\theta_s$.
A first-order approximation of $\theta_s$ was suggested in M11, but it lacked rigorous proof.

The aim of the paper is to generalise the results described in \citet{Marquet_2015_WGNE_thetas2} and \citet{Marquet_Geleyn_2015} and to derive, in sections~\ref{section_lambdas} and \ref{section_approximations}, accurate first-  and second-order approximations of $\theta_s$, written hereafter as $(\theta_s)_1$ and $(\theta_s)_2$, respectively.
These approximations are used in section~\ref{section_consequences} to compute accurate formulations for the tendencies, gradients and turbulent fluxes of moist-air entropy, with some physical properties derived from these approximations of $\theta_s$.
A conclusion is presented in section~\ref{section_conclusion}.

\section{Definition of $\theta_s$ and $(\theta_s)_1$.} 
\label{section_thetas1}
\vspace*{-3mm}

The moist-air entropy potential temperature $\theta_s$ is defined in M11 from Eq.(\ref{eq_thetas}) as the product of several terms, leading to
\vspace{-2mm}
\begin{align}
  {\theta}_{s} & \: = \; ({\theta}_{s})_1  \;
    \left( \frac{T}{T_r}\right)^{\!\!\lambda \,q_t}
    \left( \frac{p}{p_r}\right)^{\!\!-\kappa \,\delta \,q_t}
 \nonumber \\
 & \; \; \; \; \; \;
   \times \; \;
    \left( \frac{r_r}{r_v} \right)^{\!\!\gamma\,q_t}
   \frac{(1\!+\!\eta\,r_v)^{\,\kappa \, (1+\,\delta \,q_t)}}
     {(1\!+\!\eta\,r_r)^{\,\kappa \,\delta \,q_t}}
  \label{eq_thetas}
  \: ,
\end{align}
where $T$ is the temperature, $p$ the pressure, $r_v$ the water vapour mixing ratio, $q_t = q_v + q_l + q_i$ the total water specific content and $q_v$, $q_l$ and $q_i$ the water-vapour, liquid and ice specific contents.

The reference temperature $T_r$ and pressure $p_r$ are set to the standard values $T_0 =273.15$~K and $p_0=1000$~hPa in M11, where it is shown that the reference value $s_{ref} = s_{d}(T_0,p_0) - c_{pd} \: \ln(T_0)  \approx 1139$~J~K${}^{-1}$~kg${}^{-1}$, the specific value value $s$ and the potential temperature $\theta_s$ defined by Eqs.~(\ref{eq_s_thetas}) and (\ref{eq_thetas}) are all independent of any other values chosen for the reference values $T_r$ and $p_r$.
The constant moist-air reference entropy $s_{ref}$ computed with the standard values $T_0$ and $p_0$ remains unchanged for any other values of $T_r$ and $p_r$ (see Table~1 of M11).

The potential temperature
\vspace{-2mm}
\begin{equation}
    ({\theta}_{s})_1 =   \theta_{il} \;  \exp\! \left(  \Lambda_r \: q_t  \right)
  \:
  \label{eq_thetas1}
\end{equation}
that appears in Eq.~(\ref{eq_thetas}) was considered in M11 as the leading order approximation of ${\theta}_{s}$, where 
\vspace{-2mm}
\begin{equation}
 {\theta}_{il} \; = \;
 \theta \;  \exp\! \left[ \: - \:
 \frac{L_v(T)\:q_l + L_s(T)\:q_i}{{c}_{pd}\:T}
 \: \right]
  \:
  \label{eq_thetal}
\end{equation}
is close to the liquid-ice value of \citet{Tripoli_Cotton_1981} and is a generalisation of the liquid water potential temperature of \citet{Betts_73}. 
The potential temperature $\theta = T \: (p/p_0)^\kappa$ in Eq.~(\ref{eq_thetal}) is the usual dry-air version, and the latent heat of vaporization $L_v(T)$ and sublimation $L_s(T)$ depend on the absolute temperature.

The thermodynamic constants in Eqs.(\ref{eq_s_thetas})-(\ref{eq_thetal}) are those used in the ARPEGE model:
$R_d \approx 287.06$~J~K${}^{-1}$~kg${}^{-1}$, 
$R_v \approx 461.53$~J~K${}^{-1}$~kg${}^{-1}$, 
${c}_{pd}\approx 1004.7$~J~K${}^{-1}$~kg${}^{-1}$, 
${c}_{pv}\approx 1846.1$~J~K${}^{-1}$~kg${}^{-1}$, 
$\kappa = R_d/{c}_{pd} \approx 0.2857$, 
$\lambda = {c}_{pv}/{c}_{pd} - 1 \approx 0.8375$, 
$\delta = R_v/R_d - 1 \approx 0.6078$, 
$\eta = R_v/R_d \approx 1.6078$,
$\varepsilon = R_d/R_v \approx 0.622$,
$\gamma = \kappa \: \eta = R_v/{c}_{pd} \approx 0.4594$,
$L_v(T_r)=2.501\:10^6$~J~kg${}^{-1}$ and
$L_s(T_r)=2.835\:10^6$~J~kg${}^{-1}$.

The new term
\vspace{-2mm}
\begin{equation} 
\Lambda_r = \left[  \: (s_{v})_r - (s_{d})_r \: \right] \, / \, c_{pd} \; \approx \; 5.87
\label{def_Lambda_r}
\end{equation}
depends on reference specific entropies of dry air and water vapour at $T_r=273.15$~K, denoted by $(s_{d})_r = s_{d}(T_r, e_r)$ and $(s_{v})_r = s_{v}(T_r, p_r-e_r)$, where
$p_r=1000$~hPa is the reference total pressure and $e_r \approx 6.11$~hPa is the water vapour saturating pressure at $T_r$. 
The two reference entropies 
$(s_{v})_r \approx 12673 $~J~K${}^{-1}$ and 
$(s_{d})_r \approx  6777$~J~K${}^{-1}$
are computed in M11 from the third law of thermodynamics and they correspond to $(s_{v})_0 \approx 10320$~J~K${}^{-1}$ and $(s_{d})_0 \approx 6775$~J~K${}^{-1}$ computed at $T_0=273.15$~K and $p_0=1000$~hPa in HH87.
The reference mixing ratio defined in M11 by $r_r = \varepsilon \: e_r \, / \,  (p_r - e_r) \approx 3.82 $~g~kg${}^{-1}$ makes $\theta_s$ independent of $T_r$ and $p_r$.

\section{A tuning to observed and simulated datasets.} 
\label{section_lambdas}
\vspace*{-3mm}

\begin{figure}[hbt]
\centering
\includegraphics[width=0.99\linewidth]{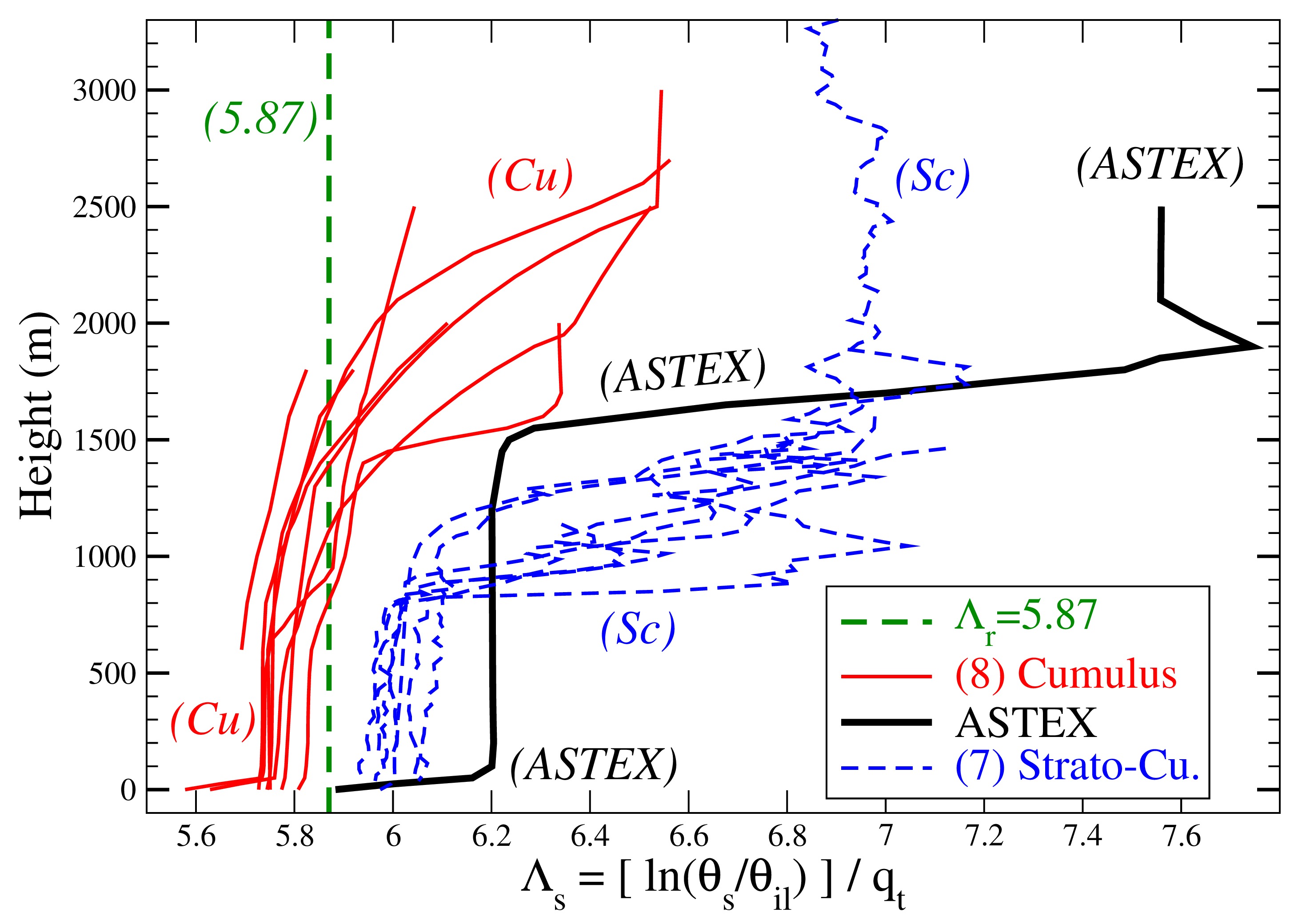}
\vspace*{-8mm} 
\caption{\small \it 
A plot of $\Lambda_s$ given by Eq.~(\ref{def_lambda_s}) for 8 cumulus (dashed blue), 7 stratocumulus (solid red) and ASTEX (solid black) vertical profiles.
The vertical green dashed line represents the value $5.87$ given by Eq.~(\ref{def_Lambda_r}).
\label{fig_Lambda}}
\end{figure}

In order to determine which factors in Eq.(\ref{eq_thetas}) may have smaller impacts (i.e. close to $1$), and to demonstrate that $({\theta}_{s})_1$ is indeed the first-order approximation of ${\theta}_{s}$, let us define the quantity $\Lambda_{s}$ by 
${\theta}_{s} = \theta_{il} \;  \exp\! \left(  \Lambda_s \: q_t  \right)$, 
where  
${\theta}_{s}$, ${\theta}_{il}$ and $q_t$ 
are known quantities and $\Lambda_s$ the unknown quantity, leading to
\vspace{-2mm}
\begin{equation} 
\Lambda_s  = \frac{1}{q_t} \; \ln\!\left(  \frac{{\theta}_{s}}{{\theta}_{il}}   \right).
\label{def_lambda_s}
\end{equation}
In order to analyse the discrepancy of $\Lambda_s$ from the constant value $\Lambda_r \approx 5.87$  given by (\ref{def_Lambda_r}), values of $\Lambda_s$ computed with Eq.~(\ref{def_lambda_s}) are plotted in Fig.~\ref{fig_Lambda} for a series of 16 observed or simulated vertical profiles of stratocumulus and cumulus.

The observed FIRE-I radial flights (02, 03, 04, 08, 10) are those studied in 
\citet{Roode_Wang_W07} and M11.
The profiles for GATE, BOMEX and ASTEX are described in \citet{Cuijpers_Bechtold_95} and those for SCMS-RF12 and DYCOMS-II-RF01 in \citet{Neggers_2003} and in \citet{Zhu_al_2005}.
The profiles for EPIC are taken from \citet{Breth_al_2005}, for ATEX from \citet{Stevens_al_2001} and for ARM-Cumulus from \citet{Lenderink_al_2004}.

The low-level values of $\Lambda_s$ remain close to the first-order value $5.87$ for the moist parts of all profiles in Fig.~\ref{fig_Lambda}, with however a standard deviation of the order of $\pm 0.2$, which may be important for certain applications.
Moreover, $\Lambda_s$ increases with height up to $6.7$ for the drier, upper-level parts of all strato-cumulus, and up to $7.6$ for the ASTEX profile.

These findings offer some insight into the way $\Lambda_s$ varies with humidity, as the more humid the profiles (low-levels and cumulus profiles), the smaller the value of $\Lambda_s$, and the drier the profiles (upper levels and strato-cumulus profiles), the larger the value of $\Lambda_s$, with ASTEX providing the driest profile.
Therefore, an accurate formulation of $\theta_s$ should be based on a increase in $\Lambda_s$ with decreasing values of water content.

\begin{figure}[hbt]
\centering
\includegraphics[width=0.99\linewidth]{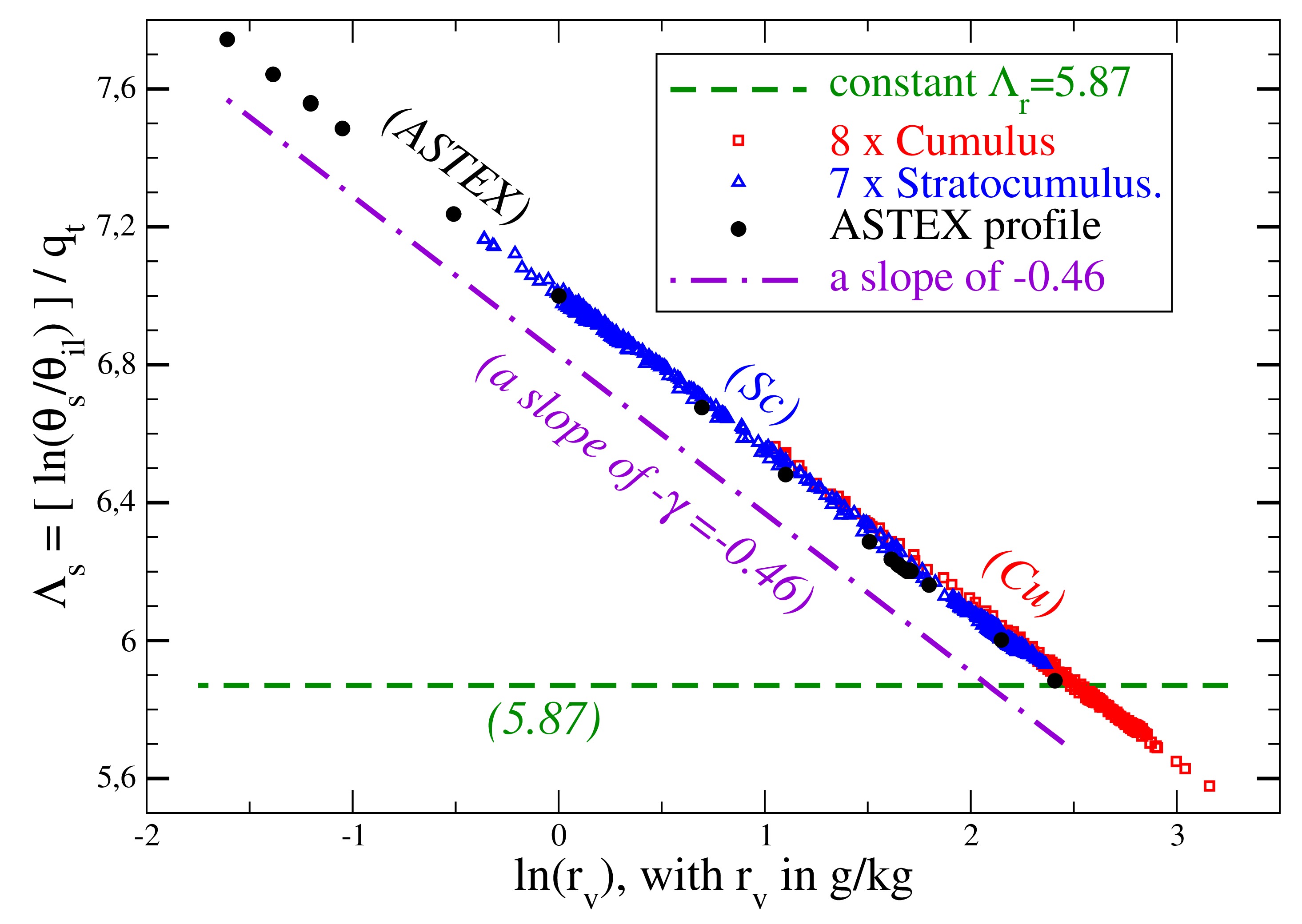}
\vspace*{-8mm} 
\caption{\small \it 
Values of $\Lambda_s$ given by Eq.~(\ref{def_lambda_s}) plotted against $\ln(r_v)$ for the same cumulus, stratocumulus and ASTEX vertical profiles as in Fig.\ref{fig_Lambda}.
The constant value $\Lambda_r \approx 5.87$ corresponds to the horizontal dashed  green line.
An arbitrary line with a slope of $-0.46$ is plotted as a dashed-dotted purple line.
\label{fig_Lambda_Lambda_s_lnrv}}
\end{figure}

A trial and error process has shown that plotting $\Lambda_s$ against $\ln(r_v)$ leads to the relevant results shown in Fig.~\ref{fig_Lambda_Lambda_s_lnrv}, where all stratocumulus and cumulus profiles are nearly aligned along the same straight line with a slope of about $- 0.46$, which may correspond to the constant $- \, \gamma$ that appears in the term $(r_v/r_r)^{- \: \gamma \: q_t}$ in Eq.~(\ref{eq_thetas}).
This very good linear fitting law appears to be valid for a large range of $r_v$ (from $0.2$ to $24$~g~kg${}^{-1}$).

It is thus useful to find a mixing ratio $r_{\ast}$ for which
\vspace{-2mm}
\begin{align} 
\Lambda_s
& \;  = \; 
\Lambda_r \: - \: \gamma \: \ln\,( r_v / r_{\ast})
\label{def_lambda_star1} \:
\end{align}
holds true, where $r_{\ast}$ will play the role of positioning the dashed-dotted line of slope $- \,\gamma \approx - 0.46$ in order to overlap the cumulus and stratocumulus symbols in Fig.~\ref{fig_Lambda_Lambda_s_lnrv}.
The unknown mixing ratio $r_{\ast}$ can be determined from Eq.~(\ref{def_lambda_star1}), rewritten as
\vspace{-2mm}
\begin{align} 
r_v  
  \; = \; 
  r_{\ast}  \;
  \exp\left( \frac{\Lambda_r \: - \: \Lambda_s }{\gamma}  \right)
\label{def_r_star1} \: ,
\end{align}
which corresponds to a linear adjustment of $r_v$ against the quantity $\exp[\: (\Lambda_r - \Lambda_s)/\gamma \:]$, where the mixing ratio $r_{\ast}$ represents the slope of the vertical profiles or scattered data points.

\begin{figure}[hbt]
\centering
\includegraphics[width=0.99\linewidth]{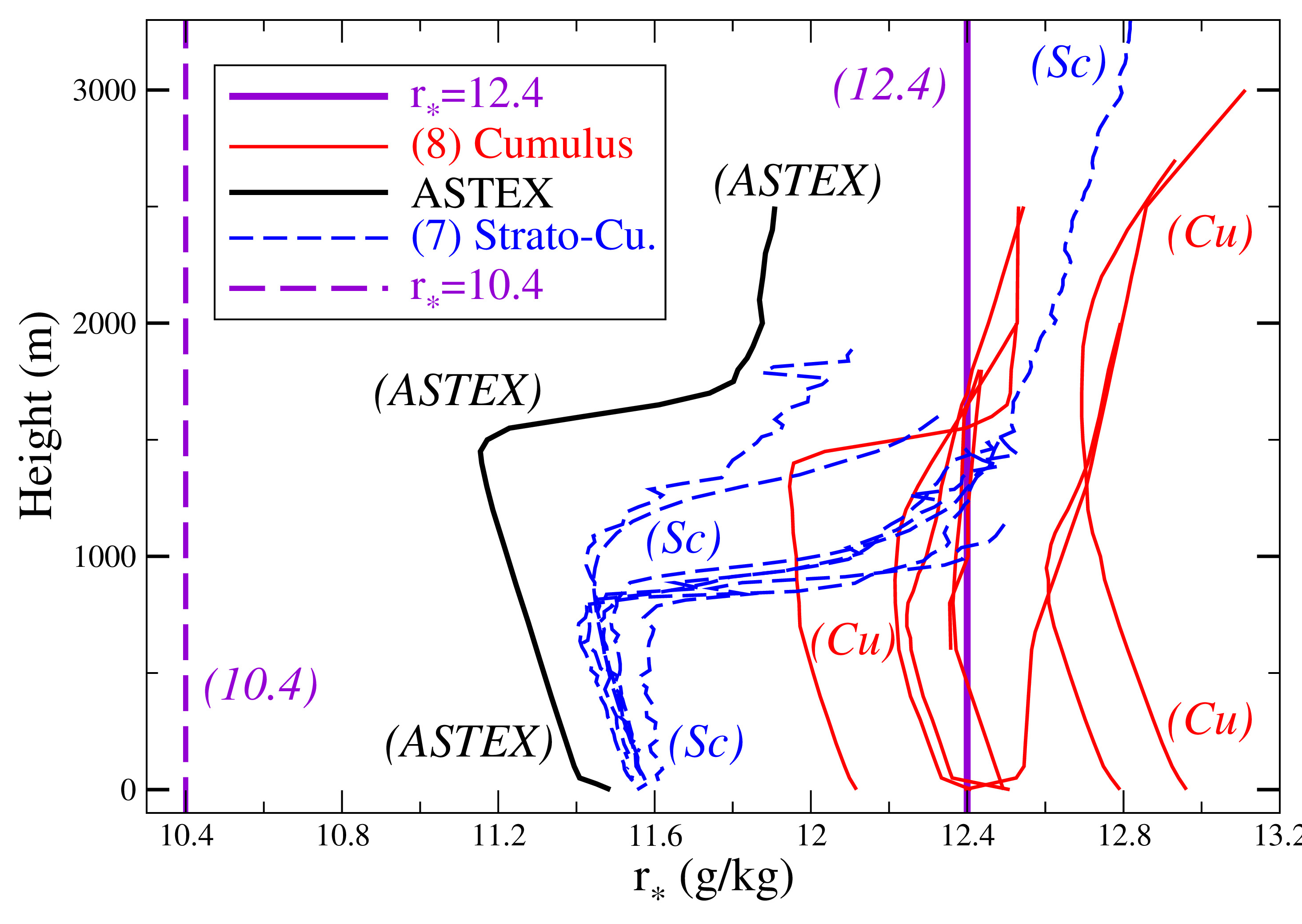}
\vspace*{-8mm} 
\caption{\small \it 
The vertical profile of $r_{\ast}(z)$ given by Eq.~(\ref{def_r_star1}) plotted for the same cumulus, stratocumulus and ASTEX vertical profiles as in Fig.\ref{fig_Lambda}.
The two vertical purple lines represent the constant values $10.4$~g~kg${}^{-1}$ and $12.4$~g~kg${}^{-1}$.
\label{fig_RVSTAR}}
\end{figure}

\begin{figure}[hbt]
\centering
\includegraphics[width=0.99\linewidth]{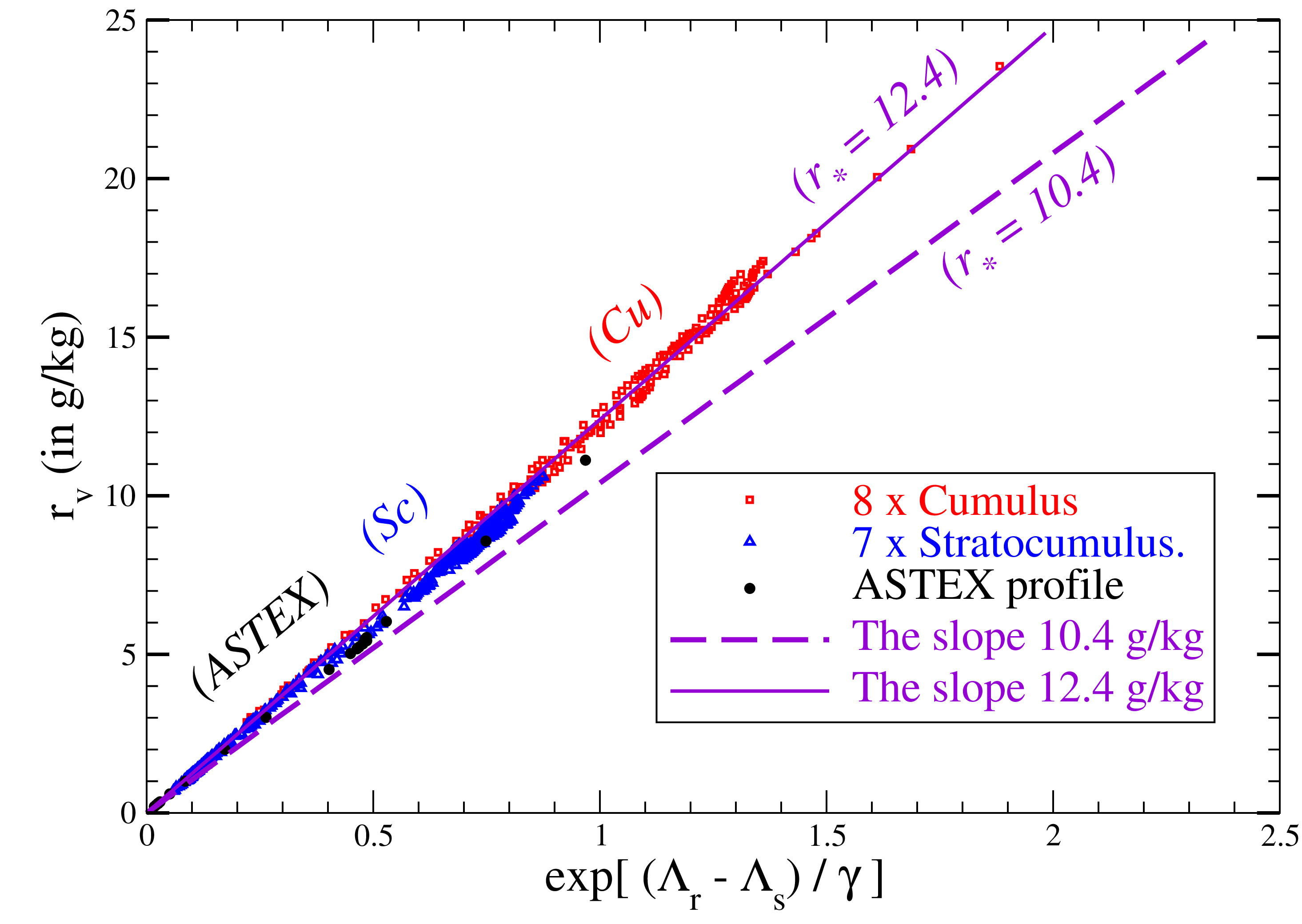}
\vspace*{-8mm} 
\caption{\small \it 
Values of $r_v$ plotted against the quantity $\exp[\: (\Lambda_r - \Lambda_s)/\gamma \:]$ according to Eq.~(\ref{def_r_star1}) and for the same cumulus, stratocumulus and ASTEX vertical profiles as in Fig.\ref{fig_Lambda}.
The two slantwise purple lines represent the special slopes of values $r_{\ast} = 10.4$~g~kg${}^{-1}$ and $12.4$~g~kg${}^{-1}$.
\label{fig_Lambda_rv_rstar}}
\end{figure}

It is shown in Figs.\ref{fig_RVSTAR} and \ref{fig_Lambda_rv_rstar} that $r_{\ast} \approx 12.4$~g~kg${}^{-1}$ corresponds to a relevant tuning of all cumulus and stratocumulus vertical profiles for a range of $r_v$ up to $24$~g~kg${}^{-1}$, whereas $r_{\ast} \approx 10.4$~g~kg${}^{-1}$ is a less relevant value introduced in the next section.

\section{Mathematical derivations of approximations of $\theta_s$.} 
\label{section_approximations}
\vspace*{-3mm}

It is possible to confirm that $ ({\theta}_{s})_1$ corresponds to the leading order approximation of ${\theta}_{s}$, and that the slope of $- \,\gamma \approx - 0.46$ with $r_{\ast} \approx12.4$~g~kg${}^{-1}$ corresponds to a relevant second order approximation for $\theta_s$, using mathematical arguments.
These results were briefly mentioned in \citet{Marquet_Geleyn_2015} and partially described in \citet{Marquet_2015_WGNE_thetas2}.
The proof is better formulated in this section and is extended to cloudy regions with liquid water or ice.

First- and second-order approximations of $\theta_s$ can be derived by computing Taylor expansions for all factors in Eq.~(\ref{eq_thetas}) for $\theta_s$, where the total water ($q_t$), the water vapour ($q_v$ and $r_v$) and the condensed water ($q_l+q_i$) specific contents or mixing ratio are considered as small quantities of the order of $1/100$ (or $10$~g~kg${}^{-1}$).

The term $(r_r/r_v)^{(\gamma\:q_t)}$ is exactly equal to the exponential $\exp[\: - \: ( \gamma \: q_t )\: \ln(r_v/r_r) \: ]$, without approximation.
The terms 
 $(T/T_r)^{\lambda \,q_t}$
and
 $(p/p_r)^{-\kappa \,\delta \,q_t}$
are similarly equal to
$\exp[\: ( \lambda \: q_t )\: \ln(T/T_r) \: ]$
and
$\exp[\: - \: ( \kappa \,\delta \: q_t )\: \ln(p/p_r) \: ]$, respectively and
without approximation.
 
The first-order expansion of  
$(1 + \eta \: r_v)^{[\: \kappa \: (1+\delta \:q_t)\: ]}
=
\exp[\: \kappa \: (1+\delta \:q_t) \: 
\ln(1 + \eta \: r_v) \: ]$ 
can be computed for small $r_v \approx q_t \approx 0.01$ 
with the help of 
$\kappa \: \eta = \gamma$,
$\ln(1 + \eta \: r_v) \approx \eta \: r_v$
and $(1+\delta \:q_t) \approx 1$, 
leading to the first-order expansion
$\exp( \gamma \: r_v )$.
Similar arguments lead to the first-order expansion 
$(1 + \eta \: r_r)^{( \kappa \: \delta \: q_t )} \approx 1$
valid for small $q_t \approx 0.01$ and $r_r \approx 0.004$.

The first-order Taylor expansion of $\theta_s$ can thus be written as
\vspace{-2mm}
\begin{align}
   & 
   {\theta}_{s}
     \approx  
  {\theta}_{il}
 \; \exp \left[ \:
    \Lambda_r \: q_t  
    \: - \:
    \gamma   \: q_t \: \ln\!\left(\frac{r_v}{r_r} \right) 
    \: + \:
    \gamma  \: r_v 
    \:\right]
\nonumber \\
   &  \; \; \; \; \; \; \; \;
 \times
 \; \exp \left[ \:
    \lambda \: q_t \: \ln\!\left(\frac{T}{T_r} \right) 
      \: - \: 
    \kappa \: \delta \: q_t \: \ln\!\left( \frac{p}{p_r} \right)
      \: \right]
\: , \label{def_THs_approx}
\end{align}
where $\theta_{il}$ is the generalized Tripoli and Cotton and Betts potential temperatures given by Eq.~(\ref{eq_thetal}).

\begin{table}
\caption{\small \it 
Values of $\ln(\theta_{\ast}/T_{\ast})$ for the OACI vertical profile and for a series of height $z$ (m) and pressure $p$ (hPa), where $\theta_{\ast}$ (K) is given by Eq.~(\ref{def_theta_star}).
The constants are: $T_{\ast}=255$~K, $p_{\ast} = 450$~hPa and $\kappa \, \delta / \lambda \approx 0.2073$.
\vspace*{2mm}
\label{Table_theta_star}}
\centering
\begin{tabular}{|c|c|c|c|c|c|}
\hline 
$z $&$p$&$T\:\mbox{(C)}$&$T\:\mbox{(K)}$&$\theta_{\ast}$&$\ln(\theta_{\ast}/T_{\ast})$ \\ 
\hline 
$10,000$ & $265$  & $-50.0$ & $223.15$ & $249.0$ & $-0.024$ \\ 
 $9,000$ & $307$  & $-43.5$ & $229.65$ & $248.6$ & $-0.025$ \\ 
 $8,000$ & $357$  & $-37.0$ & $236.15$ & $247.8$ & $-0.029$ \\ 
 $7,000$ & $411$  & $-30.5$ & $242.65$ & $247.3$ & $-0.031$ \\ 
 $6,000$ & $471$  & $-24.0$ & $249.15$ & $246.8$ & $-0.033$ \\ 
 $5,000$ & $541$  & $-17.5$ & $255.65$ & $246.1$ & $-0.036$ \\ 
 $4,000$ & $617$  & $-11.0$ & $262.15$ & $245.5$ & $-0.038$ \\ 
 $3,500$ & $658$  &  $-7.8$ & $265.35$ & $245.3$ & $-0.039$ \\ 
 $3,000$ & $700$  &  $-4.5$ & $268.65$ & $245.1$ & $-0.040$ \\ 
 $2,500$ & $746$  &  $-1.3$ & $271.85$ & $244.8$ & $-0.041$ \\ 
 $2,000$ & $794$  &   $2.0$ & $275.15$ & $244.6$ & $-0.042$ \\ 
 $1,500$ & $845$  &   $5.3$ & $278.45$ & $244.4$ & $-0.043$ \\ 
 $1,000$ & $900$  &   $8.5$ & $281.65$ & $244.0$ & $-0.044$ \\ 
   $500$ & $955$  &  $11.8$ & $284.95$ & $243.8$ & $-0.045$ \\ 
     $0$ & $1013$ &  $15.0$ & $288.15$ & $243.5$ & $-0.046$ \\ 
\hline
\end{tabular}
\end{table}

\begin{figure*}[hbt]
\centering
\includegraphics[width=0.9\linewidth]{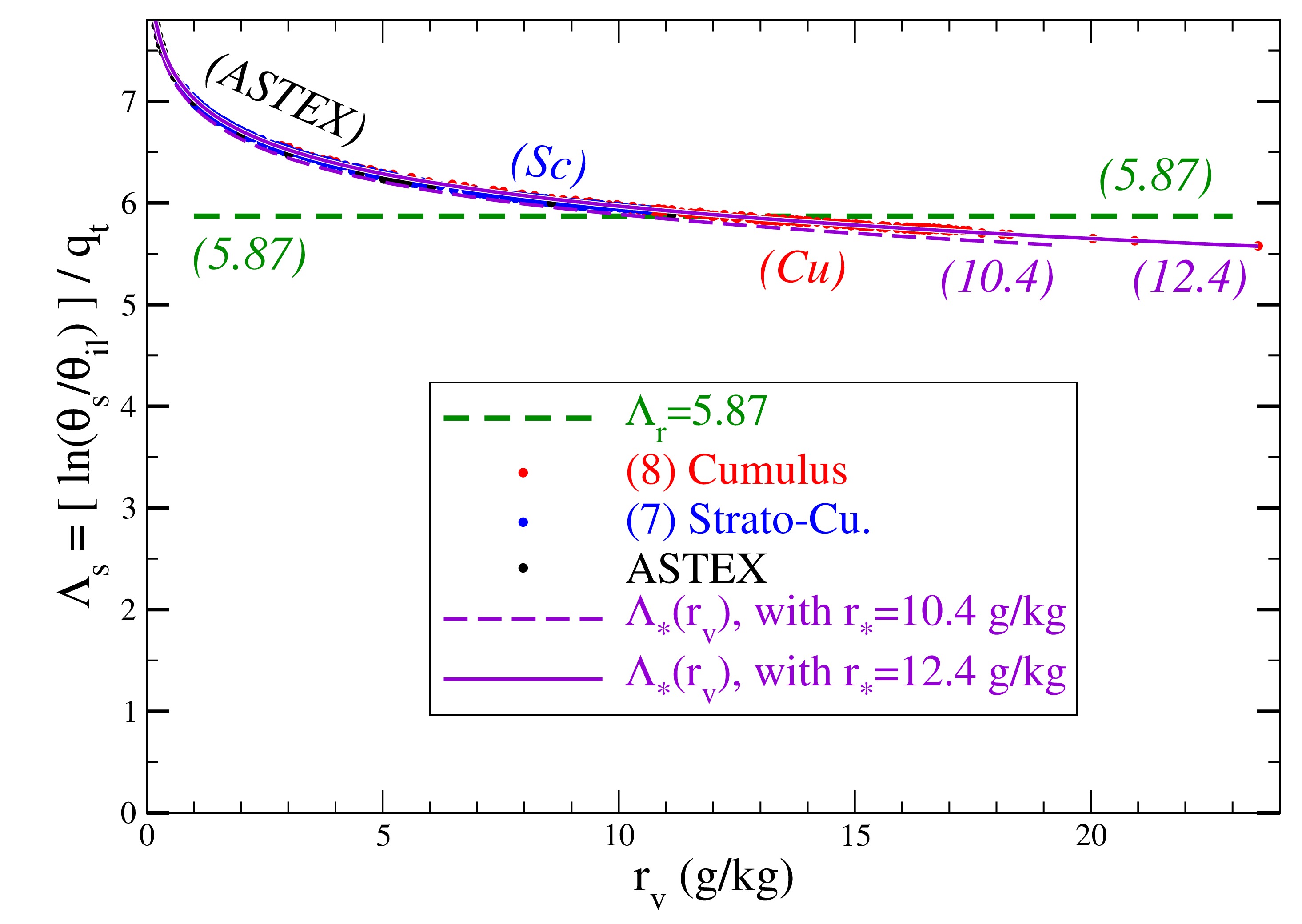}
\vspace*{-2mm} 
\caption{\small \it 
Values of $\Lambda_s$ given by Eq.~(\ref{def_lambda_s}) are plotted with symbols against the mixing ratio $r_v$ for the same cumulus, stratocumulus and ASTEX vertical profiles as in Fig.\ref{fig_Lambda}.
The two purple curves represent values of $\Lambda_{\ast}(r_v, r_{\ast}, q_l, q_i)$ given by (\ref{def_Lambda_star}) for $r_{\ast} = 10.4$ and $12.4$~g~kg${}^{-1}$.
The constant value $\Lambda_r \approx 5.87$ corresponds to the horizontal dashed green line.
\label{fig_rv_Lambda_s_rv_nozoom_points}}
\end{figure*}

The last term in the first exponential of Eq.~(\ref{def_THs_approx}) can be expressed as an equation 
\vspace{-2mm}
\begin{equation}
\gamma \: r_v \: = \: \gamma \: q_t 
\: - \: \gamma \: (q_t-q_v)
\: + \: \gamma \: q_v \: q_t / (1-q_t)
\nonumber \: ,
\end{equation}
for which the first-order approximation is obtained by dropping the last term, leading to
\vspace{-2mm}
\begin{equation}
\gamma \: r_v \: \approx \:
- \: \gamma \: q_t \: \ln[ \: 1/\exp(1) \: ]
\: - \: \gamma \: (q_l+q_i)
\nonumber \: ,
\end{equation}
where $\exp(1) \approx 2.718$ is the basis of the natural logarithms.
The second exponential of Eq.~(\ref{def_THs_approx}) can be transformed by introducing the two scaling factors $T_{\ast}$ for the absolute temperature and $p_{\ast}$ for the pressure, leading to the Taylor expansion of $\theta_s$
\vspace{-2mm}
\begin{align}
   {\theta}_{s}
     & \approx \;
  {\theta}_{il}
       \: \exp \left( 
            \Lambda_{\ast} \: q_t
                \right)
 \; \exp \left[ \:
    \lambda \: q_t \: \ln\!\left(\frac{\theta_{\ast}}{T_{\ast}} \right) 
      \: \right]
\: , \label{def_THs_approx2}
\end{align}
where
\vspace{-2mm}
\begin{align}
{\Lambda}_{\ast} & 
   \: = \; 
      \Lambda_r \: - \: 
      \gamma \; \ln\left(\frac{r_v}{r_{\ast}}\right)
   \: - \: 
      \gamma \: \left(\frac{q_l+q_i}{q_t}\right)
 \: ,  \label{def_Lambda_star}
\\
   \theta_{\ast} & 
   \:  = \;
   T \: \left(\frac{p_{\ast}}{p}\right)^{\kappa \, \delta / \lambda}
   \: , \label{def_theta_star}
\\
    r_{\ast} & \: = \;
    \: r_r
    \; \exp(1) \; 
    \left( \frac{T_{\ast}}{T_r}\right)^{\!\!\lambda / \gamma}
    \left( \frac{p_r}{p_{\ast}}\right)^{\!\!\kappa \,\delta / \gamma}
 \: . \label{def_r_star}
\end{align} 
The first two terms of $r_{\ast}$ in Eq.~(\ref{def_r_star}) represent the value $r_r \times \exp(1) \approx 10.4$~g~kg${}^{-1}$ tested for tuning the points and lines in Figs.\ref{fig_RVSTAR} and \ref{fig_Lambda_rv_rstar}.
The more accurate value $r_{\ast} \approx 12.4$~g~kg${}^{-1}$ corresponds to the mean atmospheric conditions $T_{\ast} \approx 255$~K and $p_{\ast} \approx 450$~hPa inserted into the last two terms in parentheses in Eq.~(\ref{def_r_star}).

Table~\ref{Table_theta_star} shows that the term $\ln(\theta_{\ast}/T_{\ast})$ is very small and is almost constant with height for these values of $T_{\ast}$ and $p_{\ast}$.
The term $\ln(\theta_{\ast}/T_{\ast}) \approx -0.04$ is indeed small in comparison with $\ln(r_v/r_{\ast})$, which varies between $-5$ and $+0.5$ for $r_v$ between $0.1$~g~kg${}^{-1}$ and $20$~g~kg${}^{-1}$ in the atmosphere.
It can further be show that $\ln(\theta_{\ast}/T_{\ast})$ is small by noting that $\ln(r_v/r_{\ast}) = \pm 0.04$ corresponds to values of $r_v$ within the small interval $12$ and $13$~g~kg${}^{-1}$, which is much smaller than the range of water vapour content in the atmosphere.

Similarly, the changes of $\ln(\theta_{\ast}/T_{\ast})$ in the vertical (less than $\pm 0.001$ for displacements of $500$~m) are smaller than the $10$ times larger impact of about $\pm 0.010$ for the term $(q_l+q_i)/q_t$, due to the rapid changes of typically $\pm 0.1$~g~kg${}^{-1}$ in $500$~m for $q_l+q_i$ in clouds, where $q_t \approx 10$~g~kg${}^{-1}$.

The impact of the term $\ln(\theta_{\ast}/T_{\ast})$ is thus expected to be small in comparison with the other terms, and the second exponential in Eq.~(\ref{def_THs_approx2}) can be discarded (namely, it is close to $1$ and almost constant with height).
Therefore, the relevant approximation of $\theta_s$ is made of the first two terms in the r.h.s. of Eq.~(\ref{def_THs_approx2}), leading to
\vspace{-2mm}
\begin{align}
   ({\theta}_{s})_2
    & \: =  \;
 \theta_{il} \;
       \: \exp \left( 
            \Lambda_{\ast} \: q_t
         \right)           
\: , \label{def_THs2a} \\
   ({\theta}_{s})_2
   &  \: =  \; 
  {\theta}_{il}
 \; \exp \left[ \:
    \Lambda_r \: q_t  
     - 
    \gamma \: \ln\!\left(\frac{r_v}{r_{\ast}} \right)  q_t 
     - 
    \gamma  \: (q_l+q_i) 
    \:\right]
\: , \label{def_THs2b} \\
   ({\theta}_{s})_2
    & \: =  \;
 \theta \;  \exp\! \left[ \: - \:
 \frac{L_v(T)\:q_l + L_s(T)\:q_i}{{c}_{pd}\:T}
 \: \right]
       \: \exp \left( 
            \Lambda_r \: q_t
                \right)
                 \nonumber \\
    & \; \; \; \; \times
       \: \exp \left[ \: 
    - \: \gamma \; 
    \ln\!\left(\frac{r_v}{r_{\ast}}\right)
             \: q_t
             \: \right]                
       \: \exp \left[ \: 
    - \: \gamma \: (q_l+q_i) 
             \: \right]                
\: , \label{def_THs2}
\end{align}
where $\theta_{il}$, $\Lambda_{\ast}$ and $r_{\ast} \approx 12.4$~g~kg${}^{-1}$ are given by Eqs.~(\ref{eq_thetal}), (\ref{def_Lambda_star}) and (\ref{def_r_star}), respectively.

Equations~(\ref{def_THs2a})-(\ref{def_THs2}) form a different formulation of the second-order approximation of $\theta_s$ denoted by $(\theta_s)_2$, as they include terms depending on  $\Lambda_r \: q_t \approx 0.06$ and $\gamma \: q_t \approx \gamma \: (q_l+q_i) \approx 0.005$.

In contrast, the first-order approximation is given by Eq.~(\ref{eq_thetas1}) and the first line of Eq.~(\ref{def_THs2}); i.e., by neglecting the second line composed of second order terms depending on $\gamma \: q_t$ and $\gamma \: (q_l+q_i)$, or equivalently by setting $\gamma = 0$.
This is due to the small ratio $\gamma/\Lambda_r \approx 1/13$.

Fig.\ref{fig_rv_Lambda_s_rv_nozoom_points} shows that $\Lambda_s$ defined by Eq.~(\ref{def_lambda_s}) can indeed be approximated by the second-order approximation $\Lambda_{\ast}(r_v, r_{\ast}, q_l, q_i)$ given by Eq.~(\ref{def_Lambda_star}), with  improved accuracy in comparison to the constant first-order value $\Lambda_r \approx 5.87$.
This very good tuning is valid for a range of $r_v$ between $0.2$ and $24$~g~kg${}^{-1}$.
The non-linear curves of $\Lambda_{\ast}$ with $r_{\ast} = 10.4$ or $12.4$~g~kg${}^{-1}$ both simulate the non-linear variation of $\Lambda_s$ with $r_v$ and the rapid increase of $\Lambda_s$ for $r_v < 5$~g~kg${}^{-1}$ with good accuracy.

The second exponential of Eq.(\ref{def_THs_approx2}) can be discarded (i.e., it is close to $1$) for the cumulus and strato-cumulus profiles extending up to $3$~km in Figs~\ref{fig_Lambda} and \ref{fig_RVSTAR}. 
However, this exponential may be taken into account for applications to the higher troposphere or the stratosphere regions, and especially in deep-convection clouds or in fronts where $q_t$ may be large.
For these reasons it is easy and always possible to compute and study the full version of $\theta_s$ given by Eq.(\ref{eq_thetas}), in the same way that it would be preferable to take the exact formulation of \citet{Emanuel_94} for $\theta_e$ with all exponential terms, rather than the approximate formulation of \citet{Betts_73}.

\section{Physical properties of approximations of $\theta_s$.} 
\label{section_consequences}
\vspace*{-4mm}

The tendency, vertical derivative and turbulent flux of $\theta_s$ can be evaluated by computing the differential $d\theta_s$ with the first- and second-order approximations of $\theta_s$ given by Eqs.~(\ref{eq_thetas1}) and (\ref{def_THs2a})-(\ref{def_THs2}).

\subsection{Comparisons of the third-law, equivalent and TEOS-10 entropies.} 
\label{subsection_TEOS10}
\vspace*{-3mm}

Let's analyse first the impact of the approximations of $\theta_s$ on the computations of the vertical changes of the specific moist-air entropy $s(\theta_s)$.

\begin{figure*}[hbt]
\centering
\includegraphics[width=0.99\linewidth]{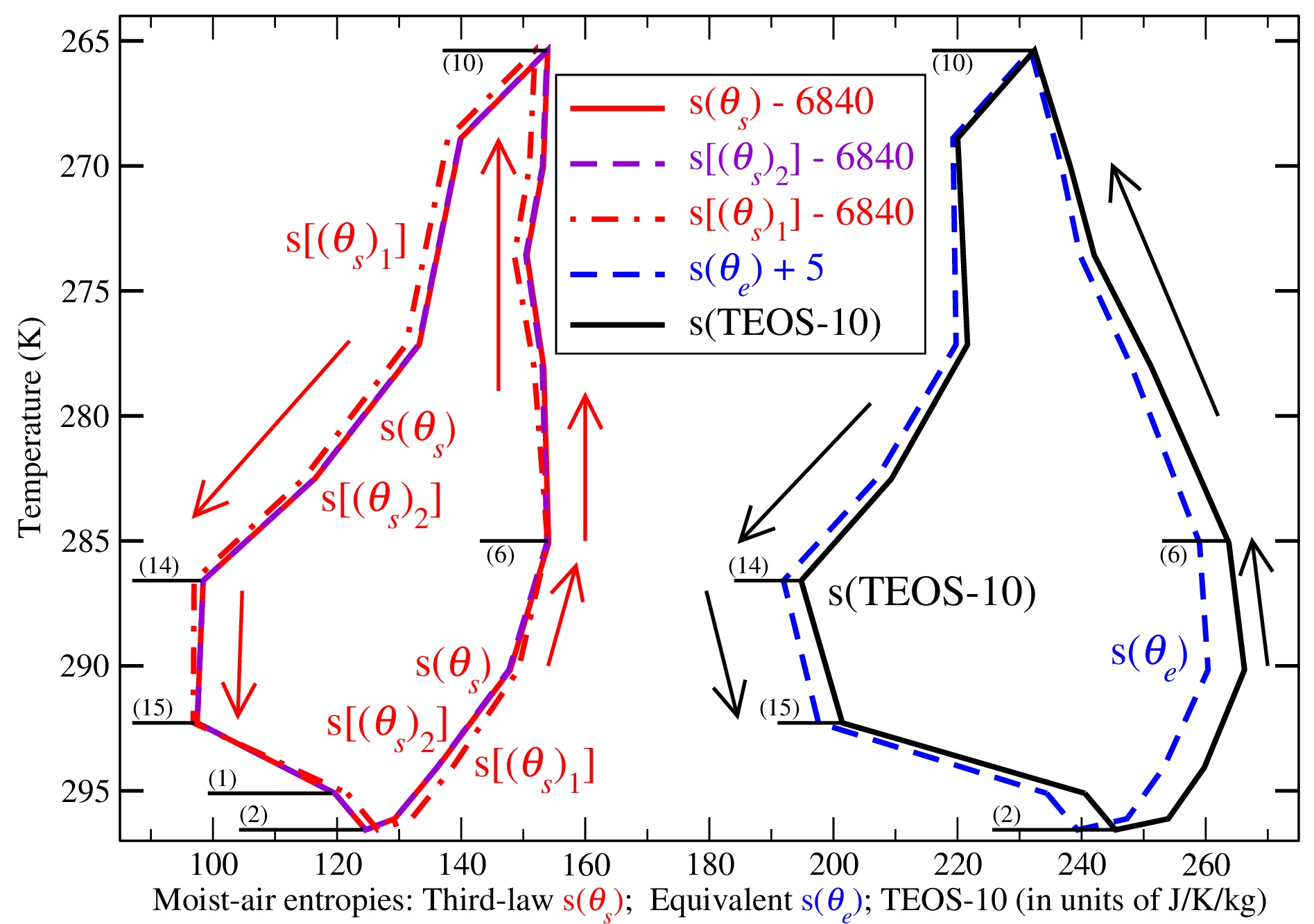}
\vspace*{-2mm} 
\caption{\small \it 
Comparisons of moist-air entropies (J~K${}^{-1}$~kg${}^{-1}$) computed for 15 points describing a loop in the Hurricane Dumil\'e simulated by the French model ALADIN:
the third-law formulations $s(\theta_s)$, $s[\,(\theta_{s})_2\,]$ and $s[\,(\theta_{s})_1\,]$ (red lines); 
an ``equivalent'' formulation $s(\theta_e)$ (blue dashed lines); 
the IAPWS-2010 and TEOS-10 formulation $s(\mbox{TEOS-10})$ (black lines).
Global offsets (the same for all points of a given loop) of $-6840$~J~K${}^{-1}$~kg${}^{-1}$ and $+5$~J~K${}^{-1}$~kg${}^{-1}$ are applied to the third-law entropy and ``equivalent'' loops, respectively. 
No offset is applied to the IAPWS-TEOS-10 version.
\label{fig_Thetas_Thetae_TEOS10}}
\end{figure*}

\begin{table}
\caption{Values of the specific entropies
$ss = s(\theta_s) - 6840$, 
$ss2 = s[\,(\theta_{s})_2\,]- 6840$,
$ss1 = s[\,(\theta_{s})_1\,]- 6840$,
$se = s(\theta_e) + 5$ and
$sT = s(\mbox{TEOS-10})$
plotted in the Fig.~\ref{fig_Thetas_Thetae_TEOS10}.
Pressure ($p$) in hPa, absolute temperature ($T$) in Kelvin, water vapour mixing ratio ($r_v$) in g~kg${}^{-1}$, entropies in J~K${}^{-1}$~kg${}^{-1}$.
\label{Table_thetas_thetae_TEOS10}}
\centering 
\begin{tabular}{ccccccccc}
\hline 
 N   &  $p$  &  $T$     &  $r_v$  & $ss$ & $ss2$ & $ss1$ & $se$ & $sT$ \\ 
\hline 
$1$ \!\!&\!\!$950$\!\!&\!\!$295.10$\!\!&\!\!$16.25$\!\!&\!\!119.6\!\!&\!\!119.6\!\!&\!\!121.7\!\!&\!\!234.4\!\!&\!\!240.5\!\!\\ 
$2$ \!\!&\!\!$950$\!\!&\!\!$296.56$\!\!&\!\!$16.24$\!\!&\!\!124.6\!\!&\!\!124.6\!\!&\!\!126.6\!\!&\!\!239.3\!\!&\!\!245.5\!\!\\ 
$3$ \!\!&\!\!$950$\!\!&\!\!$296.12$\!\!&\!\!$17.45$\!\!&\!\!129.3\!\!&\!\!129.3\!\!&\!\!132.0\!\!&\!\!247.3\!\!&\!\!254.0\!\!\\ 
$4$ \!\!&\!\!$900$\!\!&\!\!$294.07$\!\!&\!\!$17.11$\!\!&\!\!136.1\!\!&\!\!136.1\!\!&\!\!138.6\!\!&\!\!253.3\!\!&\!\!259.8\!\!\\ 
$5$ \!\!&\!\!$800$\!\!&\!\!$290.16$\!\!&\!\!$15.41$\!\!&\!\!147.9\!\!&\!\!147.7\!\!&\!\!149.2\!\!&\!\!260.4\!\!&\!\!266.3\!\!\\ 
$6$ \!\!&\!\!$700$\!\!&\!\!$285.08$\!\!&\!\!$12.64$\!\!&\!\!154.1\!\!&\!\!153.8\!\!&\!\!153.9\!\!&\!\!259.0\!\!&\!\!263.7\!\!\\ 
$7$ \!\!&\!\!$600$\!\!&\!\!$278.05$\!\!&\!\! $8.94$\!\!&\!\!153.3\!\!&\!\!153.1\!\!&\!\!151.7\!\!&\!\!248.1\!\!&\!\!251.2\!\!\\ 
$8$ \!\!&\!\!$550$\!\!&\!\!$273.57$\!\!&\!\! $6.90$\!\!&\!\!150.6\!\!&\!\!150.4\!\!&\!\!148.6\!\!&\!\!239.7\!\!&\!\!242.0\!\!\\ 
$9$ \!\!&\!\!$500$\!\!&\!\!$270.03$\!\!&\!\! $4.87$\!\!&\!\!153.2\!\!&\!\!153.1\!\!&\!\!151.0\!\!&\!\!236.8\!\!&\!\!238.2\!\!\\ 
$10$\!\!&\!\!$450$\!\!&\!\!$265.38$\!\!&\!\! $2.84$\!\!&\!\!153.9\!\!&\!\!153.8\!\!&\!\!151.9\!\!&\!\!231.8\!\!&\!\!232.4\!\!\\ 
$11$\!\!&\!\!$500$\!\!&\!\!$268.89$\!\!&\!\! $3.35$\!\!&\!\!140.0\!\!&\!\!139.9\!\!&\!\!137.9\!\!&\!\!219.3\!\!&\!\!220.0\!\!\\ 
$12$\!\!&\!\!$600$\!\!&\!\!$277.15$\!\!&\!\! $5.95$\!\!&\!\!133.2\!\!&\!\!133.1\!\!&\!\!131.1\!\!&\!\!219.8\!\!&\!\!221.6\!\!\\ 
$13$\!\!&\!\!$700$\!\!&\!\!$282.52$\!\!&\!\! $7.40$\!\!&\!\!116.4\!\!&\!\!116.3\!\!&\!\!114.6\!\!&\!\!206.9\!\!&\!\!209.4\!\!\\ 
$14$\!\!&\!\!$800$\!\!&\!\!$286.59$\!\!&\!\! $8.49$\!\!&\!\! 98.4\!\!&\!\! 98.4\!\!&\!\! 96.9\!\!&\!\!191.9\!\!&\!\!194.8\!\!\\ 
$15$\!\!&\!\!$900$\!\!&\!\!$292.28$\!\!&\!\!$10.90$\!\!&\!\! 97.4\!\!&\!\! 97.5\!\!&\!\! 96.8\!\!&\!\!197.6\!\!&\!\!201.4\!\!\\ 
\hline 
\end{tabular}
\end{table}

Fig.~\ref{fig_Thetas_Thetae_TEOS10} shows the loops for $s(\theta_s)$, $s[\,(\theta_{s})_2\,]$, $s[\,(\theta_{s})_1\,]$, $s(\theta_e)$ and $s(\mbox{TEOS-10})$ plotted for the 15 points describing a closed loop in the Hurricane Dumil\'e and published in \cite{Marquet17a}.
The pressures, temperatures and mixing ratios of these unsaturated points are listed in  Table~\ref{Table_thetas_thetae_TEOS10}.
The loop plotted with the formulation $s(\theta_e)$ of \cite{Mrowiec_al_2016} is based on an ``equivalent'' potential temperature $\theta_e$ similar to those of \citet{Betts_73} and \citet{Emanuel_94}.
The IAPWS-2010 (International Association for the Properties of Water and Steam) and TEOS-10 (Thermodynamic Equation of Seawater) formulation $s(\mbox{TEOS-10})$ is computed with the  ``SIA'' (Seawater Ice Air) software available at http://www.teos-10.org/software.htm and described in \citet{Feistel_al_2010} and \citet{Feistel_2018}.

The curves for $s(\theta_s)$ and the second-order approximation $s[\,(\theta_{s})_2\,]$ are almost superimposed, with differences of less than $0.3$~J~K${}^{-1}$~kg${}^{-1}$ according to values of $ss$ and $ss2$ in Table~\ref{Table_thetas_thetae_TEOS10}.
The differences between $s(\theta_s)$ and the first-order approximation $s[\,(\theta_{s})_1\,]$ are also small.
They are less than $1$ to $3$~J~K${}^{-1}$~kg${}^{-1}$,  which is less than one tenth of the changes of $\pm 30$~J~K${}^{-1}$~kg${}^{-1}$ in the moist-air entropy along the loop.
These results are confirmations of the good accuracy of the approximations of $\theta_s$ by $(\theta_{s})_2$ and $(\theta_{s})_1$ for various conditions of pressure, temperature and water content.

The entropy $s(\mbox{TEOS-10})$ is close to the entropy $s(\theta_e)$ of \cite{Mrowiec_al_2016}, which corresponds to the use of an equivalent potential temperature. 
This is due to the fact that the same assumptions are used to calculate the TEOS-10 and $\theta_e$ formulations: assume zero values for liquid-water and dry-air entropies at the triple point temperature of $273.16$~K.
For the same reason, the two loops for $s(\mbox{TEOS-10})$ and $s(\theta_e)$ are very different from those for $s(\theta_s)$, $s[\,(\theta_{s})_2\,]$ and $s[\,(\theta_{s})_1\,]$ since $s(\theta_s)$ is calculated with the third law, which implies the cancellation of the entropies of the most stable solid forms for all species at $0$~K.

The way the specific entropy increases or decreases with height is completely different in Fig.\ref{fig_Thetas_Thetae_TEOS10}.
The entropy changes indicated by the arrows clearly show that the variations are often of opposite signs for the TEOS-10 and third-law formulations: before point (6); and between points (14) and (15).
Moreover, the third-law values increase by about $29$~J~K${}^{-1}$~kg${}^{-1}$ between  point (2) in the boundary layer and point (10) in the middle troposphere, whereas $s(\mbox{TEOS-10})$ decreases by about $13$~J~K${}^{-1}$~kg${}^{-1}$.
Similarly, the third-law value is almost constant ($154.1$ versus $153.9$) between points (6) and (10), whereas the equivalent and TEOS-10 values decrease by about $31$~J~K${}^{-1}$~kg${}^{-1}$.

Such opposite differences of the order of $\pm 20$~J~K${}^{-1}$~kg${}^{-1}$ between vertical changes in $s(\theta_s)$, $s[\,(\theta_{s})_2\,]$ and $s[\,(\theta_{s})_1\,]$ on the one hand, $s(\theta_e)$ and $s(\mbox{TEOS-10})$ on the other hand, are large and must have significant physical impacts.
They are similar to the vertical changes in entropies shown here on Fig.\ref{fig_Thetas_Thetae_TEOS10}, and also in Figs.~19 and 20 of \citet{Feistel_al_2010}, in Figs.~1 and 2 of M11 and in Fig.~7 of \citet{Marquet17a}.

An example of such a physical impact concerns the ``heat input'' defined by the integral
$W_H = \oint T \: ds$, which is equal to the area of the loops in the ``$T-s$'' diagram shown in Fig.\ref{fig_Thetas_Thetae_TEOS10}.
It is one part of the work received by a parcel of moist-air undergoing a closed loop.
The area $W_H$ is about $34$~\% larger with $s(\theta_e)$ and $s(\mbox{TEOS-10})$ than with the third-law value $s(\theta_s)$.
These large differences are similar to those published in \cite{Marquet17a} and they are bound to have an important physical meaning for convection considered as a thermal machine, because the impact on $W_H$ of the choice of the reference state for entropies is not balanced by the impact of the other part of $W$ depending on the water content \citep{Marquet17a}.

Moreover, since the entropy is a state function, it cannot decrease or increase at the same time between two points, depending on the choice of the references values for the entropies of liquid water and dry air, and $W_H$ cannot have an indeterminate value depending on these references values.
Otherwise, this would contradict the second law itself, because one could create or destroy entropy at will just by changing the reference values.

Since the arbitrary choices retained in TEOS-10 and in ``equivalent'' formulations may have an impact on atmospheric energetics, the only relevant choice is the third-law definition given by \cite{Planck_1917} and retained in HH87 and M11.
The same reasons impose to use the third-law definition of the entropies for all species in order to analyse the stability of chemical reactions.
Therefore, it would be interesting to modify the TEOS-10 definitions and computations by taking into account the third-law values for entropies, which are available in HH87 and M11 for dry air and liquid water and in Thermodynamical and Chemical Tables for all atmospheric species.

\subsection{The differentials of $\theta_s$.} 
\label{subsection_consequences_diff}
\vspace*{-3mm}

The differential of $(\theta_s)_2$ is computed from Eq.~(\ref{def_THs2b}), leading to
\begin{align}
\vspace{-2mm}
\frac{d(\theta_s)_2}{(\theta_s)_2} 
& \: = \;
\frac{d\theta_{il}}{\theta_{il}} 
\: + \:  
 \left[ \: 
 \Lambda_r
 - \gamma \: \ln\!\left(\frac{r_v}{r_{\ast}}\right)
 \: \right] 
\: dq_t
\nonumber \\
& \; \; \; \: - \: 
 \left[ \: 
\gamma
\;
\frac{q_t}{r_v}
 \: \right] 
\: dr_v
\: - \:  
 \left[ \: 
    \gamma 
 \: \right] 
\: d q_c
\: , \nonumber
\end{align}
where $r_v=(q_t-q_c)/(1-q_t)$ depends on $q_t$ and $q_c=q_l+q_i$.
This differential of $(\theta_s)_2$ can thus be written in terms of $d\theta_{il}$, $dq_t$ and $dq_c$, leading to
\vspace{-2mm}
\begin{align}
\hspace*{-3mm}
\frac{d(\theta_s)_2}{(\theta_s)_2}  
& \: = \:
\frac{d\theta_{il}}{\theta_{il}} 
\: + \:  
A_{t} \: dq_t
\: + \:  
A_{c} \: dq_c
\: ,
\label{eq_dths2a} \\
\hspace*{-3mm}
d(\theta_s)_2 
& \: = \:
A_{\theta} \; d\theta_{il}
\: + \:  
A_{t} \: (\theta_s)_2 \: dq_t
\: + \:  
A_{c} \: (\theta_s)_2 \: dq_c
\: ,
\label{eq_dths2}
\end{align}
where
\vspace{-2mm}
\begin{align}
A_{\theta} & = \: 
\exp({\Lambda}_{\ast}\:q_t)
\; , \label{A_theta}  \\
A_{t} & = 
 \left[ \: 
 \Lambda_r
 - \gamma \: \ln\!\left(\frac{r_v}{r_{\ast}}\right)
 - \gamma \:
     \left(
     \frac{q_t}{q_v}
     \right)
     \left(
     \frac{1-q_c}{1-q_t}
     \right)
 \: \right]
\; , \label{A_qt}  \\
A_{c} & = \:
\gamma
 \: \left(
\frac{r_c}{r_v}
     \right)
  \; = \:
\gamma
 \: \left(
\frac{q_c}{q_v}
     \right)
\; . \label{A_qil} 
\end{align}

The first-order approximation is obtained by setting $\gamma = 0$ in Eqs.(\ref{def_Lambda_star}) and (\ref{A_theta})-(\ref{A_qil}), leading to
\vspace{-2mm}
\begin{align}
d(\theta_s)_1 
& \: = \:
\exp({\Lambda}_r\:q_t) \; d\theta_{il}
\: + \:  
 \Lambda_r
 \:
 (\theta_s)_1 
 \: dq_t
\: .
\label{eq_dths1} 
\end{align}

Moreover, the first-order approximations of the moist-air entropy ($\theta_s$) and Betts  potential temperatures ($\theta_l$ and $\theta_e$) can be further simplified and compared with the crude assumptions $q_i=0$, $\Lambda_r \approx 6$ and $L_v \approx 9 \: c_{pd} \: T$, leading to
\vspace{-2mm}
\begin{align}
\hspace*{-3mm}
 \theta_l \:
&
 \: \approx \; 
 \theta \:
 \exp\left( 
      - \: 9 \: q_l
 \right)
\nonumber\: , \\
\hspace*{-3mm}
 \theta_e
&
 \: \approx \; 
 \theta \:
 \exp\left( 
      + \: 9 \: q_v
 \right)
 \: \approx \; 
 \theta_l \:
 \exp\left( 
      + \: 9 \: q_t
 \right)
\nonumber \: , \\
\hspace*{-3mm}
 \theta_s
 \; \approx \; 
(\theta_s)_1
&
 \: \approx \; 
 \theta_l \:
 \exp\left( 
      + \: 6 \: q_t
 \right)
 \: \approx \; 
 \theta_e \:
 \exp\left( 
      - \: 3 \: q_t
 \right)
\label{eq_approx_ths1} \: .
\end{align}

\subsection{The tendencies of $\theta_s$.} 
\label{subsection_consequences_tendencies}
\vspace*{-3mm}

The differentials given by Eqs.~(\ref{eq_dths2}) and (\ref{eq_dths1}) can be used to compute the tendencies ($d \psi/dt$ or $\partial \psi/\partial t$) for any scalar variable $\psi$, leading for instance to the time derivative of the first-order moist-air entropy potential temperature
\vspace{-2mm}
\begin{align}
\!\!\!
\frac{d(\theta_s)_1}{dt} 
& \: = \:
\exp({\Lambda}_r\:q_t) \; \frac{d \theta_{il}}{dt} 
\: + \:  
\Lambda_r \: \theta_{il} \: \exp({\Lambda}_r\:q_t) \:\frac{d q_t}{dt} 
\: .
\label{eq_dths1dt} 
\end{align}
According to Eq.~(\ref{eq_approx_ths1}), the tendency of the specific moist-air entropy can thus be approximated by
\vspace{-2mm}
\begin{align}
\frac{ds}{dt}
& \: = \; 
 \frac{c_{pd}}{\theta_s} \: \frac{d\theta_s}{dt}
  \: \approx \; 
 \frac{c_{pd}}{({\theta}_{s})_1} \: \frac{d({\theta}_{s})_1}{dt} 
\nonumber \: , \\
\frac{ds}{dt}
& \: \approx \; 
 \frac{c_{pd}}{\theta_l} 
 \left( \: \frac{d\theta_l}{dt} 
 \: + \:  6 \; \theta_l \: \frac{dq_t}{dt} 
 \right)
\label{eq_dthetal_dt} \: , \\
\frac{ds}{dt}
& \: \approx \; 
 \frac{c_{pd}}{\theta_e} 
 \left( \: \frac{d\theta_e}{dt} 
 \: - \: 3 \;\theta_e \: \frac{dq_t}{dt} 
 \right)
\label{eq_dthetae_dt} \: .
\end{align}

The impacts on entropy changes of the terms ${dq_t}/{dt}$ in Eqs.~(\ref{eq_dthetal_dt}) and (\ref{eq_dthetae_dt}) can be similar or larger than those of  ${d\theta_l}/{dt}$ and ${d\theta_e}/{dt}$, because $6 \: \theta_l$ and $3 \: \theta_e$ are of the order of $1800$ and $1000$, respectively.
Therefore the change in entropy due to an increase in $\theta_l$ or in $\theta_e$ of about $1$~K can be balanced by the impact of a decrease in $q_t$ of about $0.6$ or $1$~g~kg${}^{-1}$.
Values of this order of magnitude were obtained for the ``diabatic forcing'' evaluated by \cite{Yanai_al_1973} and \cite{Johnson_al_2016} in studies of deep convection, where the vertical profiles of apparent heat sources and moisture sinks leads to values at $900$~hPa close to 
$d\theta/dt \approx Q_1/c_{pd} \: \approx +1$ to $+1.5$~K~day${}^{-1}$ and $dq_v/dt \approx - \: Q_2/L_v \: \approx - 0.8$ to $-1.2$~g~kg${}^{-1}$~day${}^{-1}$, respectively.
These values lead to almost no entropy changes and may correspond to the constant moist-air entropy regime described in M11 in the boundary layer of marine strato-cumulus.

These findings prove that the change in the moist-air specific entropy must be computed by employing $\theta_s$, or its first- or second-order approximations $({\theta}_{s})_1$ or $({\theta}_{s})_2$, and cannot be computed by using changes in the Betts variables $\theta_l$ or $\theta_e$ alone.
The terms ${dq_t}/{dt}$ in Eqs.~(\ref{eq_dthetal_dt}) and (\ref{eq_dthetae_dt}) must be taken into account with those factors close to $+6$ and $-3$ corresponding to the third-law definition of the specific entropies of dry air and water vapour.

\subsection{The diabatic changes of $\theta_s$.} 
\label{subsection_consequences_diabatic}
\vspace*{-3mm}

The ``diabatic'' heating rate is usually computed from the total derivative $d\theta/dt$.
It is assumed that the dry-air potential temperature $\theta$ is a function of the dry-air specific entropy alone, and is thus conserved by fluid parcels when the motion is ``adiabatic''.
The heating rate $Q$ is defined by writing the equation $ds/dt = (c_{pd}/\theta) \:\:d\theta/dt = Q/T$, where $s$ is the dry air entropy.

In contrast, the study of the third-law entropy given by Eq.~(\ref{eq_s_thetas}) and of the moist-air entropy equation $ds/dt = (c_{pd}/\theta_s) \:d\theta_s/dt$, may justify replacing $\theta$ by $\theta_s$, with another definition for the ``diabatic'' heating rate $Q_s$.
The change in the first-order specific moist-air entropy 
$ds/dt \approx [ \: c_{pd}/(\theta_s)_1 \: ] \: d(\theta_s)_1/dt$ 
can be computed from Eq.~(\ref{eq_dths1dt}) with $\theta_{il}$ given by Eq.~(\ref{eq_thetal}) and by assuming the first-order hypotheses $d/dT(L_v/T) \approx d/dT(L_s/T) \approx 0$, leading to
\vspace{-2mm}
\begin{align}
\!\!\!
\frac{ds}{dt} 
& 
\: \approx \:
\frac{c_{pd}}{\theta} 
\frac{d\theta}{dt} 
 - 
\frac{L_v}{T} \: \frac{d q_l}{dt} 
 - 
\frac{L_s}{T} \: \frac{d q_i}{dt} 
 +  
c_{pd} \: \Lambda_r \: \frac{d q_t}{dt} 
\: \approx \:  
\frac{Q_s}{T}
\: .
\label{eq_ds_dtheta} 
\end{align}
The factors ${c_{pd}}/{\theta} \approx 3$, ${L_v}/{T} \approx {L_s}/{T} \approx 9000$ and $c_{pd} \: \Lambda_r \approx 6000$ explain that changes of about $1$~g~kg${}^{-1}$ due to ${d q_l}/{dt}$, ${d q_i}/{dt}$ or ${d q_t}/{dt}$ in Eq.~(\ref{eq_ds_dtheta}) lead to the same impact as a change of about $2$~K due to ${d \theta}/{dt}$.
The change of the moist-air entropy evaluated with $d\theta_s/dt$ can therefore be of a sign opposite to that of $d\theta/dt$, depending on the impacts of the changes in $q_l$, $q_i$ or $q_t$.

The main difference between the moist-air entropy Eq.~(\ref{eq_ds_dtheta}) for $\theta_s$ and the equation for $\theta$ is the conservative feature valid for $d\theta_{il}/dt$, which corresponds to an equilibrium between the three terms depending on ${d \theta}/{dt}$, $dq_l/dt$ and $dq_i/dt$.
This means that reversible phase changes have no impact on $\theta_{il}$, $\theta_s$ and the specific moist-air entropy, whereas they are interpreted as diabatic sources for $\theta$.
The other difference is the impact of entrainment, detrainment, diffusion, precipitation and evaporation processes in the atmosphere considered as an open system, because all these processes modify the specific moist-air entropy and $\theta_s$ via the change in total vapour contents  $dq_t/dt$ in Eq.~(\ref{eq_ds_dtheta}).

The apparent diabatic heating rate $Q$ acting on $T$ or $\theta$ depends on both the impact of radiation and phase changes.
Conversely, the diabatic heating rate $Q_s$ acting in the specific energy ($h - R\: T = h - p/\rho$), specific enthalpy ($h$) and specific entropy ($s$ or $\theta_s$) equations is mainly due to the impact of radiation, with no impact from reversible changes of phases.

\subsection{Links between entropy and moist static energies (MSE).} 
\label{subsection_consequences_MSE}
\vspace*{-3mm}

It is shown in \citet{Marquet17a} and \citet{Marquet_Thibaut_2018} that the slopes of the isentopes labelled with the third-law potential temperature $\theta_s$ are different from the slopes of surfaces of equal values of $\theta$, $\theta_l$, $\theta_e$ or $\theta'_w$.

Similarly, it is shown in this section that the changes in moist-air entropy and $\theta_s$ may be different from those of the sum $h + \phi$ of the potential energy $\phi=g \: z$ and the moist-air enthalpy, where $h$ is defined in
\citet{Marquet_2015_QJ_h,Marquet_2015_WGNE_h_flux} by
\vspace{-2mm}
\begin{align}
h
& 
\: = \:
c_{pd} \: T
\: - \: L_v \: q_l 
\: - \: L_s \: q_i 
\: + \: L_h \: q_t 
\: + \: h_{\rm ref}
\: 
\label{eq_h_plus_phi}
\end{align}
or equivalently, with $L_f=L_s-L_v$, by
\vspace{-2mm}
\begin{align}
h
& 
\: = \:
c_{pd} \: T
\: + \: L_v \: q_v 
\: - \: L_f \: q_i 
\: + \: (L_h-L_v) \: q_t 
\: + \: h_{\rm ref}
\: ,
\label{eq_h_plus_phib} \\
h
& 
\: = \:
c_{pd} \: T
\: + \: L_s \: q_v 
\: + \: L_f \: q_l 
\: - \: (L_s-L_h) \: q_t 
\: + \: h_{\rm ref}
\: .
\label{eq_h_plus_phic}
\end{align}
The reference constant value $h_{\rm ref} \approx 256$~kJ~kg${}^{-1}$, together with the latent heat $L_h(T) = h_v(T) - h_d(T)$, are computed in
\citet{Marquet_2015_QJ_h,Marquet_2015_WGNE_h_flux},
where it is shown that
$L_h(T) \approx 2.603\:10^6$~J~kg${}^{-1} + (c_{pv}-c_{pd}) \: (T-273.15 \,\mbox{K})$.

The sum of $\phi$ plus $h$ given by Eqs.~(\ref{eq_h_plus_phib}) or (\ref{eq_h_plus_phic}) is thus similar to the frozen moist static energy 
$\mbox{FMSE} = c_{pd} \: T + L_v \: q_v - L_f \: q_i + \phi$ 
studied in \citet{Siebesma_al_2003} and \citet{Rooy_al_13}, or to the liquid moist static energy 
$\mbox{LMSE} = c_{pd} \: T + L_s \: q_v + L_f \: q_l + \phi$ 
studied in \citet{Dauhut_al_2017}, provided that $q_t$ is a constant with $dq_t/dt=0$, or if the additional terms $(L_h-L_v) \: q_t$ or $- \: (L_s-L_h) \: q_t$ are discarded.

However, these additional terms may have significant impacts on values of $h$ if $q_t$ is not a constant, because
$L_h-L_v \approx 0.2\:10^6$~J~kg${}^{-1}$ and
$L_s-L_h \approx 0.3\:10^6$~J~kg${}^{-1}$, which are of the same order of magnitude as the latent heat of fusion $L_f \approx 0.33\:10^6$~J~kg${}^{-1}$.
This means that a change of $1$~g~kg${}^{-1}$ for $q_t$ has the same impact on the moist-air enthalpy $h$ as a change of $0.3$~K for $T$ in the atmosphere considered as an open system, namely due to entrainment, detrainment, diffusion, evaporation at the surface and precipitation processes, which all modify the dry-air and total water vapour contents, namely with $dq_t/dt \neq 0$.

The same impacts can be evaluated by computing both the differential of $s$ and of $h+\phi$, with $h$ given by any of Eqs.~(\ref{eq_h_plus_phi})-(\ref{eq_h_plus_phic}), leading to the exact formula
\vspace{-2mm}
\begin{align}
\!\!\!
d(h+\phi)
& 
\: = \:
c_p \: dT
 -  L_v \: dq_l 
 -  L_s \: dq_i 
 +  L_h \: dq_t 
 +  g \: dz
\: ,
\label{eq_d_h_plus_phi} 
\end{align}
where 
$c_p = (1-q_t) \: c_{pd} + q_v \: c_{pv} 
+ q_l \: c_l + q_i \: c_i$ is the moist-air value of the specific heat at constant pressure.
The Gibbs equation written in Eq.(16) in \citet{deGroot_Mazur_86} provides the general link between the changes in moist-air entropy $s$ and enthalpy $h$, yielding
\vspace{-2mm}
\begin{align}
T \: \frac{ds}{dt}
& 
\: = \:
\left( 
   \frac{dh}{dt} - \frac{1}{\rho} \frac{dp}{dt} 
\right)
\: - \:
\left[ \: 
\sum_{k=0}^3 \mu_k \: \frac{dq_k}{dt} 
\: \right]
\: .
\label{eq_Gibbs} 
\end{align}

The first-order approximation of $ds/dt$ given by Eq.~(\ref{eq_ds_dtheta}) can be used to evaluate the bracketed terms in Eq.~(\ref{eq_Gibbs}), namely the opposite of the sum of the Gibbs potentials $\mu_k = h_k - T \: s_k$ and the change in specific contents ${dq_k}/{dt}$ (this sum is for $k=0,1,2,3$ for dry-air, water-vapour, liquid-water and ice, respectively).
Both Eq.~(\ref{eq_d_h_plus_phi}) and the differential of the dry-air potential temperature
$d\theta/\theta = dT/T - (R_d/c_{pd})\: dp/p$ can be
inserted in Eqs.~(\ref{eq_ds_dtheta}) and (\ref{eq_Gibbs}) with $p=\rho \: R \: T$,  $c_p \approx c_{pd}$ and $R \approx R_d$, leading to the first-order approximate Gibbs entropy equation
\vspace{-1mm}
\begin{align}
T \: \frac{ds}{dt}
& 
 \approx 
\frac{c_{pd} \: T}{(\theta_s)_1} \: \frac{d(\theta_s)_1}{dt}
\nonumber \\
T \: \frac{ds}{dt}
& 
\: \approx \:
 \frac{d(h+\phi)}{dt}
- 
\left[ \: 
\left( 
   L_h - c_{pd} \: T \: \Lambda_r
\right)
\frac{dq_t}{dt} 
\: \right]
\nonumber \\
& \; \; \; \; \;
\: - \:
\left( 
   g \: \frac{dz}{dt} + \frac{1}{\rho} \frac{dp}{dt}
\right)
\: .
\label{eq_Gibbs2} 
\end{align}
The terms in parentheses in the second line of Eq.(\ref{eq_Gibbs2}) cancel out for vertical and hydrostatic motions only, namely if $dp/dt = - \: \rho \: g \: dz/dt$.
This is a first limitation for a possible link between $T\:ds$ and $d(h+\phi)$, which cannot be valid for non-hydrostatic or slantwise or horizontal motions.

Moreover, the bracketed term must be taken into account in the atmosphere considered as an open system where $dq_t/dt \neq 0$ due to irreversible diffusion, evaporating or precipitating processes.
Indeed, the factor $L_h - c_{pd} \: T \: \Lambda_r \approx 0.3\:10^6$~J~kg${}^{-1}$ is of the same order of magnitude as the latent heat of fusion $L_f$, and a change of $1$~g~kg${}^{-1}$ for $q_t$ has the same impact on the Gibbs equation as a change of $0.3$~K for the moist-air entropy potential temperature $(\theta_s)_1$. 
This means that $h+\phi$ or the MSE quantities fail to represent the changes in specific moist-air entropy for the atmosphere considered as an open system.

\subsection{The turbulent fluxes of $\theta_s$.} 
\label{subsection_consequences_turbulence}
\vspace*{-2mm}

It is explained in \citet{Richardson_19a,Richardson_19b} and \citet[p.66-68]{Richardson_22} that the moist-air turbulence must be applied to the components of the wind ($u$, $v$), the total water content $q_t$ and either the specific moist-air entropy ($s$) or the corresponding potential temperature (i.e. the third-law value $\theta_s$ derived in M11 that Richardson was not able to compute in 1922).

Accordingly, the thermodynamic variables on which the turbulence is acting in almost all present atmospheric parameterizations are the two Betts variables ($\theta_l$, $q_t$), with $\theta_l$ considered as synonymous with the specific moist-air entropy.
However, many hypotheses are made in \citet{Betts_73} to compute $\theta_l$ (and $\theta_e$) from a certain moist-air entropy equation: this is valid if and only if $R/c_p \approx R_d/c_{pd}$, $L_v(T)/T$ and $q_t$ are all assumed to be constant.
Therefore $\theta_l$ is an approximation of the moist-air entropy and is not completely determined, because any arbitrary unknown function of $q_t$  can be added or put into a factor of $\theta_l$ and $\theta_e$ in Betts formulas, with $\theta_e$ indeed derived from $\theta_l$ in \citet{Betts_73} by a mere multiplication by the arbitrary factor $\exp[\: (L_v \: q_t)/(c_{pd} \: T)\: ]$.

The third-law formulation $\theta_s$ solve these issues, and the term $\exp(\Lambda_r \: q_t)$ is one of the unknown functions of $q_t$ that was lacking in the computation of $\theta_l$ in \citet{Betts_73} as well as in \citet{Emanuel_94}, where the reference entropies are arbitrary chosen to set $\Lambda_r \approx 0$ for deriving $\theta_l$, or $\Lambda_r \approx L_v/(c_{pd} \: T ) \approx 9$ for deriving $\theta_e$, two terms which are different from the third-law value $\Lambda_r \approx 6$.

The first-order vertical turbulent flux of the third-law moist-air entropy $\theta_s$ is obtained by using the differential given by Eq.~(\ref{eq_dths1}), leading to
\vspace{-2mm}
\begin{align}
\overline{w'(\theta_s)'_1}
& \: = \:
\exp({\Lambda}_r\:q_t) \; \overline{ w' \theta'_{il} }
\: + \:  
\Lambda_r \: \theta_{il} \: \exp({\Lambda}_r\:q_t) \: \overline{w' q'_t}
\: .
\nonumber 
\end{align}
According to Eq.~(\ref{eq_approx_ths1}), the turbulent flux $\overline{w'(\theta_s)'_1}$ can then be approximated by
\vspace{-2mm}
\begin{align}
\overline{w' s'}
& \: = \; 
 \frac{c_{pd}}{\theta_s} \; \:
  \overline{w' \theta_s'}
  \: \approx \;
 \frac{c_{pd}}{(\theta_s)_1} \; \:
  \overline{w'(\theta_s)'_1}
\label{eq_w_s} \: , \\
\overline{w'(\theta_s)'_1}
& \: \approx \; 
 \frac{(\theta_s)_1}{\theta_l} 
 \left( \: \overline{w' \theta_l'}
 \: + \:  6 \; \theta_l \: \overline{w' q'_t} 
 \right)
\label{eq_w_thetal} \: , \\
\overline{w'(\theta_s)'_1}
& \: \approx \; 
 \frac{(\theta_s)_1}{\theta_e} 
 \left( \: \overline{w' \theta_e'}
 \: - \: 3 \;\theta_e \:  \overline{w' q'_t} 
 \right)
\label{eq_w_thetae} \: .
\end{align}
The physical meaning of the third-law term $\exp(\Lambda_r \: q_t)\approx 6$ in  Eqs.~(\ref{eq_w_thetal})-(\ref{eq_w_thetae}) is clear: this term precisely takes into account the impacts of $\overline{w' q'_t}$ in the atmosphere considered as an open system where the dry-air and water vapour contents $q_d=1-q_t$ are not constant.
The impacts of $\overline{w' q'_t}$ in Eqs.~(\ref{eq_w_thetal}) and (\ref{eq_w_thetae}) may be large due to the factors 
$6 \: \theta_l \approx 1800$~K and $3 \: \theta_e \approx 1000$~K.
The turbulent flux $\overline{w'(\theta_s)'_1}$ can therefore have the opposite sign to $\overline{w' \theta_l'}$, depending on the value of the flux $\overline{w' q'_t}$,  leading to possible counter-gradient terms which can be computed by Eq~(\ref{eq_w_thetal}) for the specific moist-air entropy flux that is approximately equal to $c_{pd}/(\theta_s)_1$ times $\overline{w'(\theta_s)'_1}$.

The need described by Richardson to use the third-law value $\theta_s$ for computing turbulent fluxes with $\exp(\Lambda_r \: q_t)$ and $\Lambda_r \approx 6$, and to use any of Eqs.~(\ref{eq_w_s})-(\ref{eq_w_thetae}), is confirmed by the study in M11 of the FIRE-I radial-flights 02, 03, 04, 08 and 10, where it is shown that only $\theta_s$ is well-mixed and constant in the whole boundary layer, including the entrainment region, and with almost no jump at the interface between the the boundary layer and the dry-air region above.

A corollary of the use of the specific moist-air entropy, and thus $\theta_s$ or $({\theta}_{s})_1$ or $({\theta}_{s})_2$, in the parameterizations of turbulence is described in \citet[p.177, chapter 8/2/18]{Richardson_22} prophetic book:
``although the (exchange) coefficient is provisionally taken as the same for both the entropy and the total water content, yet we must expect a discrimination between the two cases as more knowledge is gained''.
Recent results described in \citet{Marquet_Belamari_2017_WGNE_Lewis} and \citet{Marquet_al_2017_WGNE_Lewis} confirm Richardson's vision by showing that the entropy Lewis number is different from unity for the M\'et\'eopole-Flux (M\'et\'eo-France), Cabauw (KNMI), and ALBATROS terrestrial and marine datasets.

The physical consequences can be understood by computing the first-order turbulent fluxes of the dry-air and virtual potential temperatures $\theta$ and $\theta_v$ from those of $\theta_s$ and $q_t$.
The simple case of clear-air conditions ($q_l=q_i=0$ and $q_t=q_v$) is considered here, leading to
\vspace{-2mm}
\begin{align}
\!\!\!
\overline{w' \theta'_s} & \approx 
- \: K_s \; \frac{\partial \overline{\theta_s}}{\partial z}
\label{eq_w_theta_s_Ks} \: , \\
\!\!\!
\overline{w' q'_v} & \approx 
- \: K_q \; \frac{\partial \overline{q_v}}{\partial z}
\label{eq_w_theta_s_Qt} \: , \\
\!\!\!
\overline{w' \theta'} & \approx 
-  K_q \: \mbox{Le}{}_{\rm ts} \:
   \frac{\partial \overline{\theta}}{\partial z}
-  K_q \: 
  \Lambda_r \; \overline{\theta} \;
  \left( \mbox{Le}{}_{\rm ts} - 1 \right) 
   \frac{\partial \overline{q_v}}{\partial z}
\label{eq_w_theta} \: , \\
\!\!\!
\overline{w' \theta'_v} & \approx 
- \: K_q \: \mbox{Le}{}_{\rm ts} \:
   \frac{\partial \overline{\theta_v}}{\partial z}
- K_q \: 
  (\Lambda_r-\delta) \; \overline{\theta} \;
  \left( \mbox{Le}{}_{\rm ts} - 1 \right) 
   \frac{\partial \overline{q_v}}{\partial z}
\label{eq_w_thetav} \: .
\end{align}
Equations~(\ref{eq_w_theta_s_Ks}) and (\ref{eq_w_theta_s_Qt}) express the K-gradient hypothesis applied to the moist-air entropy and water content, where $K_s$ and $K_q$ are the exchange coefficients suggested by Richardson.
Equation~(\ref{eq_w_theta}) explains that the first-order turbulent flux of the Betts liquid-water potential temperature ($\theta_l = \theta$) is not proportional to ${\partial \overline{\theta}}/{\partial z}$ for the general atmospheric conditions, except for the special case $\mbox{Le}{}_{\rm ts} = K_s / K_q = 1$.
Similarly, the buoyancy flux $(g/\theta) \: \overline{w' \theta'_v}$ can be computed with Eq.~(\ref{eq_w_thetav}) and is proportional to the vertical gradient of $\theta_v$ only if $\mbox{Le}{}_{\rm ts} = 1$.

The signs of the additional terms in Eqs.~(\ref{eq_w_theta})-(\ref{eq_w_thetav}) depend on the signs of both ${\partial \overline{q_v}}/{\partial z}$ and $(\mbox{Le}{}_{\rm ts} -1)$, and since $\Lambda_r \; \overline{\theta} \approx 
(\Lambda_r - \delta) \; \overline{\theta}
\approx 1800$~K are large, the terms in the r.h.s. of Eqs.~(\ref{eq_w_theta})-(\ref{eq_w_thetav}) are of the same order of magnitude if $\mbox{Le}{}_{\rm ts} \neq 1$.
These new additional terms proportional to ${\partial \overline{q_v}}/{\partial z}$ may lead to important physical impacts in the parameterization of atmospheric turbulence, since they can act as significant direct- or counter-gradient terms.
Moreover, the limit of the small value of $\mbox{Le}{}_{\rm ts} \approx 0$ studied in \citet{Marquet_2017_WGNE_Lewis_prod} and observed in stable conditions (at night) leads to the turbulent flux
$\overline{w' \theta'} \approx 
\left[ \: K_q \: \Lambda_r \; \overline{\theta} \: \right] \;
 {\partial \overline{q_v}}/{\partial z}$
 and 
$\overline{w' \theta_v} \approx 
\left[ \: K_q \: (\Lambda_r-\delta) \; \overline{\theta} \: \right] \;
 {\partial \overline{q_v}}/{\partial z}$, 
which depends only on the vertical gradient of $q_v$.

The modified turbulent flux of $\theta_v$ given by Eq.(\ref{eq_w_thetav}) acts in the equation of turbulent kinetic energy, which can be greatly modified if $\mbox{Le}{}_{\rm ts}$ is different from unity, as it seems to happen in both stable and  unstable cases.
This may lead to new paradigms for computing and understanding the flux Richardson number and the thermal production of turbulent kinetic energy in these stable and unstable regimes where $\mbox{Le}{}_{\rm ts} \neq 1$.
Therefore, a promising application of the representation of the specific moist-air entropy by $\theta_s$, $({\theta}_{s})_2$ or $({\theta}_{s})_1$ is  the possibility to parametrize the turbulence of moist air by first calculating the fluxes of $\theta_s$ and $q_t$, to deduce that of $\theta_{il}$, with a counter-gradient term depending at the same time on the flux of $q_t$ and $\mbox{Le}{}_{\rm ts} \neq 1$.
These aspects related to the turbulence of moist air will be addressed in a paper to come.

\section{Conclusions.} 
\label{section_conclusion}
\vspace*{-4mm}

The first- and second-order approximations $({\theta}_{s})_1$ and $({\theta}_{s})_2$ of the specific moist-air entropy potential temperature $\theta_s$ are derived by using both tuning processes and mathematical arguments.
It is confirmed that $\theta_s$ can be understood as a generalisation of the two Betts variables $(\theta_l, q_t)$, with the dependence in $q_t$ of the specific moist-air entropy that could not be derived by \citet{Betts_73} and \citet{Emanuel_94} because the hypotheses $dq_t = 0$ or $q_t = \: $constant were assumed.

The first-order tendencies and vertical turbulent fluxes of $({\theta}_{s})_1$ are compared to those of the first-order approximations of the Betts variables $\theta_l$ and $\theta_e$.
It is explained that the impact of the total water content $q_t$ is large and prevents the use of $\theta_l$ and $\theta_e$ to describe or parameterize the moist-air turbulence if the entropy Lewis number is different from unity.
It should be noted that the problems posed by the multiple and very imprecise definitions of $\theta_e$ \citep[up to $3$~K or more, see][]{Marquet_2011,Marquet17a,Marquet_Thibaut_2018} are much larger than those discussed here for small differences of less than $0.6$~K between $\theta_s$ and $({\theta}_{s})_1$, and of less than $0.1$~K between $\theta_s$ and $({\theta}_{s})_2$.

More general versions of Eqs.(\ref{eq_thetas1}) and (\ref{def_THs2b}) for $(\theta_s)_1$ and $(\theta_s)_2$ can be considered by a multiplication by the factors in the third line of Eq.~(6) in \citet{Marquet17a}, namely if the mixed-phase conditions and non-equilibrium processes need to be taken into account
\citep{Marquet_2016_WGNE_mixed_phase}.
These factors concern, for instance, under- or supersaturation with respect to liquid water or ice, and/or temperature of rain or snow different from $T$.

An open question is whether is is necessary to include the precipitating species (rain, snow, graupels, ...) in $q_l$ and $q_i$ to compute $\theta_s$.
This question is addressed in \citet{Marquet_Thibaut_2018} for the very-deep convection regime of Hector the Convector, with large simulated impacts in the computation of the entropy stream-function if precipitating species are taken into account.



\vspace*{4mm}
\noindent{\bf \Large Acknowledgements}
\label{Acknowledgements}

The author want to thank the editor and the two reviewers for their comments, which
helped to improve the manuscript.
\vspace*{4mm}



\bibliographystyle{ametsoc2014}
\bibliography{Marquet_MWR_Thetas2_R1}

\end{document}